# Optical properties of circular Bragg gratings with labyrinth geometry to enable electrical contacts

*Quirin Buchinger*[*1], *Simon Betzold*[1], *Sven Höfling*[1], *and Tobias Huber-Loyola*[1]

[1]Lehrstuhl für Technische Physik, Physikalisches Institut, Julius-Maximilians-Universität Würzburg, Am Hubland, 97074 Würzburg, Germany

*quirin.buchinger@uni-wuerzburg.de

## Abstract

We present an optical study of various device designs for electrically contactable circular Bragg grating cavities in labyrinth geometries. To create an electrical connection between the central disk and the surrounding membrane, which are separated through air gaps, we introduce connections between the adjacent rings. We propose to rotate these connections, creating a labyrinth like structure, to disable waveguiding and keep the mode confinement. To investigate how different arrangement and size of the connections affect the optical properties and to find the optimal design, six different layouts with either 3-fold or 4-fold symmetry and one with 2-fold symmetry are investigated experimentally and by numerical simulations. Reflectivity measurements and simulations show that rotating the connections improves the mode confinement, far-field pattern and Purcell factor compared to layouts with connections arranged in straight lines. We compare results between different layouts for different connection widths and perform polarization resolved measurements to investigate whether the connections create asymmetries in the photonic confinement that would impede the performance of the device.

Semiconductor quantum dots (QDs) are promising sources for single or entangled photons for use in quantum cryptography [1–3] and optical quantum computing [4]. QDs benefit from high brightness and indistinguishability [5,6], when embedded in microcavities. Recently, the circular Bragg grating (CBG) resonator [7] has gained a lot of attention as a cavity to embed semiconductor quantum dots [8–12]. This attention is warranted, first, by simplified and shortened epitaxial growth, compared with resonators that rely on vertical distributed Bragg reflectors, such as micropillar resonators [5,6], and second, by high Purcell enhancement in a broad frequency range. This enables Purcell enhancement of both the biexciton and exciton emission simultaneously [11,12], without complicated coupled cavity designs as previously required [13]. However, various applications, like a blinking-free operation [14–16], deterministic charging [17] or switching of quantum dot molecules [18], require embedding the quantum emitters in a diode structure with electrical contacts on the device. Early device geometries, which were not completely etched [9], would allow for a bottom contact, but a top contact still seems out of reach. Furthermore, these partially etched devices come along with a compromise in device performance, compared to devices on oxide with fully etched rings [9,11,12]. The attempt to leave narrow waveguide-like connections is a possibility, which was already reported [19]. However, in Ref. [19] the authors report a very large device with quality (Q)-factors of ≈ 1000, to minimize the effect of the connection of the circular rings, which is contradicting the broadband enhancement mentioned above. Also, the connection must be extremely narrow to not create a waveguide towards the side. Quantum dot emission coupling to such waveguides is lost for off-chip experiments. On the other hand,

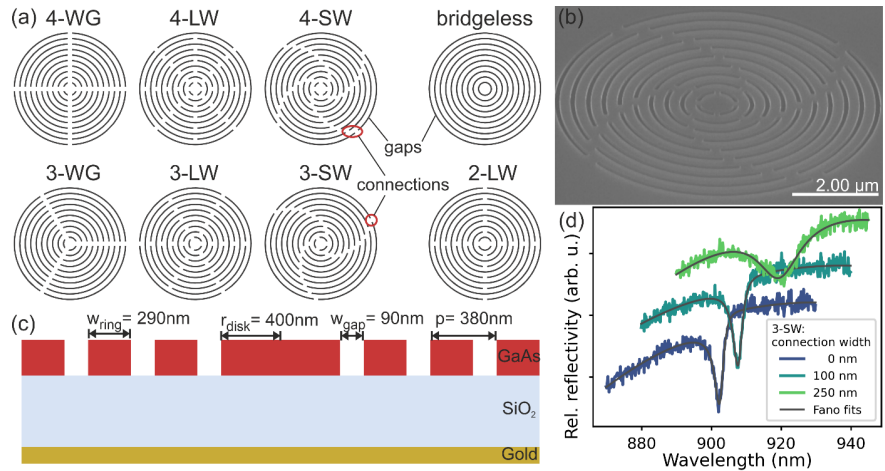

*Figure 1 a) The different layouts investigated in this work. The layout names begin with the number of connections per individual ring. Thus, the layouts 4-WG, 4-LW, 4-SW (abbreviations for "waveguide", "long way", "short way", resp.) have 4 connections in each gap and are distinct only by the shift of the connections' positions between the nearest rings. 4-WG and 3-WG layouts feature the arrangement of the connections along straight lines, while for 4-LW, 3-LW and 2-LW the connections are rotated 45°, 60° and 90°, respectively, between the consecutive gaps. The small rotation in the 3-SW and 4-SW layouts ensures a shorter electrical path between the center and the rim of CBG. Here, the layouts are shown with a connection size of 300 nm. The black features are etched into the membrane, defining the CBG structure. b) SEM picture of CBR with 4-SW layout. c) Sketch of CBG cross-section on top of $SiO_2$-layer and backside gold mirror. d) Reflectivity spectra of 3-SW CBGs with 0 nm, 100 nm and 250 nm connections plotted with an offset in y.*

the narrowness of the connections is limited, since for too small connections the electrical contact deteriorates.

Our approach to overcome this challenge is based on the intentional shifting of the position of the connections between consecutive rings. We investigate seven different layouts for the connection positions, which have 2-fold, 3-fold or 4-fold geometry as shown and named in Figure 1 (a). These layouts are compared to a bridgeless circular Bragg grating. The 4-**w**ave**g**uide (4-WG) and 3-**w**ave**g**uide (3-WG) layouts have four and three equidistant connections per etched gap, respectively, arranged in a straight line, connecting all rings in a waveguide-like fashion (for very thin connections, there is no existing waveguiding mode). The layouts labelled as 4-LW (**l**ong **w**ay) and 3-LW in Figure 1 (a) feature connections rotated 45°(4-LW) and 60°(3-LW) from one gap to the next, respectively. This results in a labyrinth-like pattern and into a long way between the membrane and the central disk and thus an increased resistance, when electrically contacting the sample. To reduce this way in the layouts 4-SW (**s**hort **w**ay) and 3-SW, see Figure 1 (b), the connections are rotated 45°(4-SW) and 60°(3-SW) only for the two innermost gaps and afterwards by 15° (8° from the 5$^{th}$ ring outwards) from one gap to the next. A scanning electron microscope (SEM) image of the 4-SW layout, forming a windmill-like pattern, is shown in Figure 1 (b). Additionally, a 2-fold geometry is fabricated, with two opposite connections per ring rotated 90° at every second gap (2-LW). For each layout, we varied the nominal width of the connection from 50 nm over 75, 100, 150, 200, 250, 300, 350, 400 and 450 nm. In the actual devices, the connection width is (28 ± 3) nm smaller than the nominal width due to process inaccuracies. All following numbers state the nominal width. At 50 nm connection width, the process did etch partly into the top of the connection.

For all the CBGs we used a (120 ± 5) nm thick GaAs membrane on top of (360 ± 10) nm of $SiO_2$ and a gold bottom mirror. The sample was transferred on a GaAs substrate via a flip chip process [10–12].

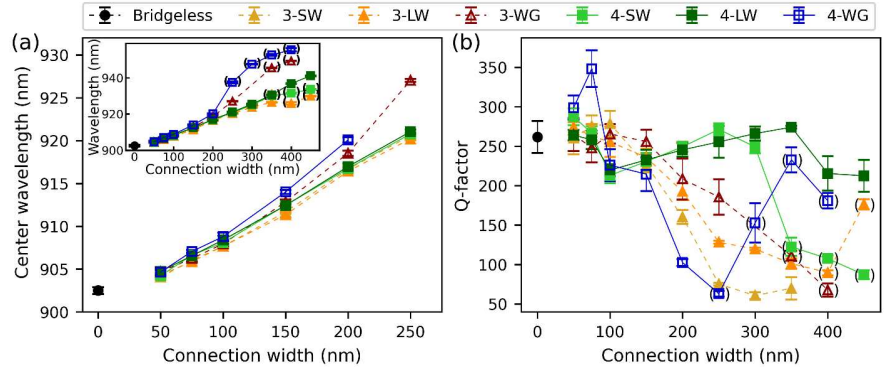

*Figure 2: Effect of the connection width on (a) center wavelength and (b) Q-factor for different layouts. In (a) the data is displayed up to 250 nm connection width for clarity. The entire data range is shown in the Inset. Each datapoint is the mean of at least 5 measured devices with the standard deviation as error bars. The excluded datapoints (marked with parentheses) represent modes which could not be clearly identified as the investigated fundamental mode (at bigger connections sometimes additional modes arise). The reduced effect of the connections on the Q-factor for the 4-LW and 4-SW layouts is demonstrated. The lines serve as a guideline for the eyes.*

We patterned the CBGs with electron beam lithography and dry etching. The CBGs have a central disk radius of 400 nm with a periodicity of 380 nm and a gap width of 90 nm (Figure 1 c).

To investigate the CBG fundamental mode and the influence of the different layouts and connection widths, we performed reflectivity measurements on the devices. Therefore, a white light source was focused on the cooled down (~4K) devices, using a 50x, NA=0.65 objective. The reflected light was channeled into a monochromator and spectrally analyzed. The acquired spectra are normalized on a reference spectrum from a silver mirror to get relative reflectivity spectra. At the cavity resonance, the spectrally flat white light is coupled into the structure resulting in a drop in reflectivity (Figure 1 d). The resulting shape is given by a Fano resonance [20], which is caused by interference of the 0-dimensional CBG mode and the 2-dimensional modes of the membrane [21]. Therefore, we fit the spectra with a Fano line shape [21,22]:

$$I(\lambda) = A \frac{(p\Gamma/2 + \lambda - \lambda_{center})^2}{(\Gamma/2)^2 + (\lambda - \lambda_{center})^2} \ , \quad (1)$$

with amplitude A, center wavelength $\lambda_{center}$, linewidth $\Gamma$ and Fano parameter p. At p=0, the Fano resonance converges to a Lorentzian function. When fitting, we take a quadratic background into account, which comes from interference of planar surfaces in the structure, i.e., the upper and lower surface of the membrane. We calculate the cavity Q-factor $Q = \lambda_{center}/\text{FWHM}$. The FWHM is given by the full width at half the intensity between the base line and the minimum of the resonance peak from the Fano fit.

For the bridgeless CBG, the cavity resonance is determined to be (902.5 ± 0.4) nm (@4 K) with a quality factor of Q=260 ± 20 (averaged over 15 measured devices). For every different device design (different layout or different connection width), we measured at least 5 devices, while we measured up to 8 devices for some of the designs. In total 301 devices were evaluated. We calculate $\lambda_{center}$ and the Q-factor from the fitted data. Figure 2 (a) displays the center wavelength of the CBG mode in dependence of the connection size for the different layouts. For all layouts, a redshift is visible with increasing connection width. There are two processes that contribute to this redshift. First, the cavity mode is less confined in the inner disk due to the presence of the connections. Second, the effective refractive index in the structure is increased because there is more high index material present. For layouts with the same number of connections, 4-WG, 4-LW,4-SW and 3-WG, 3-LW, 3-SW, respectively, this effect is expected to be the same. Since the LW and SW layouts show a reduced redshift compared to the WG layouts, they have a better confinement. This is also supported by FDTD simulations, discussed further down.

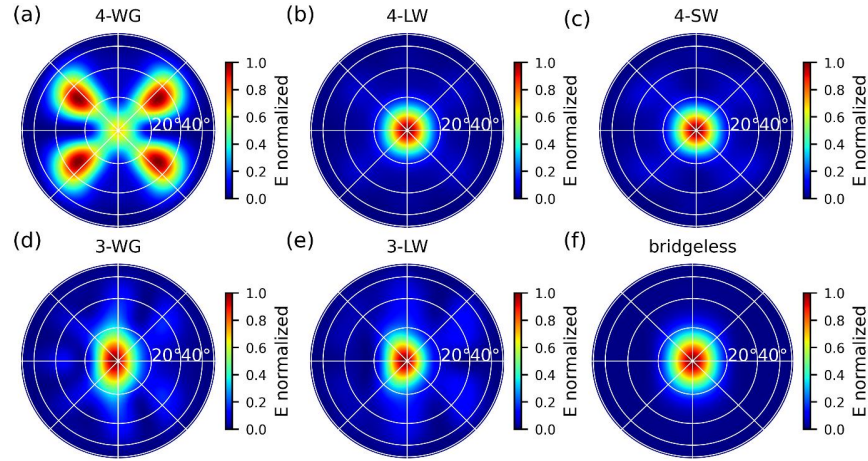

Figure 3: a)-f) Simulated far-field pattern for the different layouts with a connection width of 250 nm and a horizontal oriented dipole. Displayed is the electric field rather than the intensity for better visibility of the small deviations. The Gaussian shape of the far-field pattern of the bridgeless layout (f) is lost for 4-WG (a), 3-WG (d) and 3-LW (e). In contrast to them, the 4-LW (b) and 4-SW (c) restores the Gaussian shape.

Figure 2 (b) shows the connection width dependence of the Q-factor. For connection sizes from 0 to 100 nm, the Q-factor for all layouts stays the same within its fluctuations. Afterwards it decreases for all layouts except 4-SW and 4-LW. For these two layouts, changing the connection arrangement from WG to LW (SW) prevents the Q-factor from dropping. The Q-factor stays at the same level until 350 nm (300 nm). For all layouts with three connections per ring, a decrease of the Q-factor after 150 nm is observed, but the Q- factor is still higher than in the 4-WG layout.

Finite Difference Time Domain (FDTD)-simulations for the different designs corroborate the results. We calculate Purcell factors $F_p$, far-field pattern (Figure 3 a-f) and total emission efficiencies into the upper hemisphere as well as efficiencies into a given numerical aperture (Table 1). For the bridgeless CBG the simulations yield a Purcell factor of Fp = 19 at 932 nm whereby 89% (thereof 91% into NA=0.64) of the light is emitted upwards. For the 4-WG (250 nm connection width) the Purcell factor decreases to $F_p$ = 13 (79% light upwards, thereof 62 to 59% into NA=0.64, depending on the dipole orientation) and more importantly, the far-field pattern completely loses its Gaussian shape (Figure 3 a). For this layout the mode is oriented along the waveguide like connections and thus has some momentum towards these directions. Transformed into the far-field pattern this results in the shown pattern. The 4-LW and 4-SW layouts mitigates this issue, by design. Thus, the light is confined more homogeneous in all directions, and we get a Gaussian like far-field pattern (Figure 3 b, c). Moreover, these layouts restore the Purcell factor to to $F_p$ = 18 and 17, respectively. For the 3-WG the Purcell factor is even lower with a value of $F_p$ = 9. Rotating the connections to 3-LW and 3-SW restores the Purcell factor partially to $F_p$ = 10. In all 3-fold layouts, the far-field patterns also strongly deviate from a Gaussian distribution, see (Figure 3 d, e). The fact that the Purcell factors of the LW and SW layouts, respectively, are greater than WG layouts (Table 1) supports that LW and SW layouts have a better confinement compared to the WG layouts. Efficient fiber coupling is very important for quantum technological applications [23]. To indicate how well the emitted light can be collimated into a fiber due to the different far-field patterns, we calculate their overlap with a two dimensional (2D)-Gaussian mode profile. For this, we fit the far-field patterns with 2D Gaussian profiles and calculate the fraction of the volume of the residuals compared to the volume of the far-field pattern. We use a circular 2D-Gaussian to avoid the need for correcting for ellipticity when fiber coupling the emission. This could in principle be included and would make the overlap with the bridgeless layout slightly better. However, for a quantum dot as emitter, the two dipole directions are orthogonal, making such a correction impossible, even in principle. Thus, we decided to calculate the overlap with a symmetric Gaussian mode. The calculated values show that the overlap of the 4-LW and 4-SW layouts are in the same region as for the bridgeless CBG while values for 3-WG, 3-LW, 3-SW and especially 4-WG are lower. When discussing the extraction efficiencies, we want to point out that the CBG parameters were not optimized for the new cavity wavelength. We assume that the extraction efficiencies could be increased, because we also see a non-optimal extraction efficiency for bridgeless CBGs with non-optimized parameters.

| Layout | $F_P$ | UP | NA=0.64 | Overlap |
|---|---|---|---|---|
| bridgeless | 19 | 89% | 91% | 87% |
| 4-WG | 13-14 | 79% | 59-62% | 1-5% |
| 4-LW | 18 | 72-73% | 83-81% | 76-94% |
| 4-SW | 17 | 76-78% | 77-75% | 72-92% |
| 3-WG | 9 | 46-47% | 75-76% | 57-63% |
| 3-LW | 10 | 60-62% | 68% | 73-71% |
| 3-SW | 10 | 68% | 68% | 52-54% |

*Table 1: Simulated Purcell factor FP and the total emission efficiencies into the upper hemisphere (UP) and the fraction thereof into an NA=0.64. Calculated overlap between the far-field profiles and fitted, circular 2D-Gaussian profiles as indicator how well the emission can be coupled into single mode fibres. The given value ranges represent different simulations with different dipole orientations.*

In a perfectly round CBG the fundamental mode consists of two perpendicularly polarized degenerate modes. This degeneracy can be lifted by introducing a different confinement along different axes, as already shown in intentionally elliptical devices [24]. To investigate the effect of the introduced connections on the confinement along different directions, we performed polarization resolved reflectivity measurements. We measured three devices for each layout with connection sizes of 75 nm, 150 nm, 250 nm, and 300 nm. As a reference, nine bridgeless devices were measured. The spectra are fitted with a Fano line shape (eq. 1). For the polarization mode splitting, we evaluated the minima of the Fano resonances, because the center wavelength $\lambda_{center}$ depends on the Fano parameter p and thus does not represent the splitting correctly.

For the bridgeless CBG, one would expect zero mode splitting. However, we do observe a finite splitting in most of the devices ((0.5 ± 0.3) nm on average). Figure 4 (a) displays reflectivity measurements along the horizontal (H) and vertical (V) polarization of a bridgeless CBG, where H and V are chosen such that they coincide with one of the non-degenerate eigenmodes, respectively. The inset of Figure 4 (a) shows the related wavelengths of the minima of the fitted spectra, depending on the polarization direction. The extrema in the sinusoidal fit represent the H- and V-polarized mode positions and are used to calculate the splitting. There are three potential reasons for the non-zero splitting: unwanted elongation during the lithography (astigmatism in the writing e-beam) or anisotropic etching, which leads to a slightly elliptically CBG or anisotropic strain introduced during the flip-chip process which leads to an anisotropic refractive index.

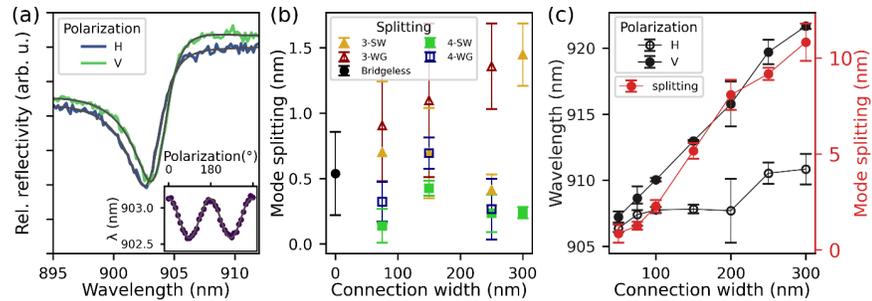

*Figure 4: a) Reflectivity spectra at H and V polarization of a bridgeless CBG. A polarization mode splitting of (0.541±0.009) nm is visible. Inset) The spectral position of the Fano resonance minima for different polarization angle. b) Polarization mode splitting for the different CBG layouts in dependence of the connection width. For layouts with three connections per ring the splitting is increased. The 3-LW and 4-LW follow basically the trend of 3-SW and 4-SW, respectively. We decided to not show these datapoints for the sake of readability of the plot. c) Polarization mode splitting and mode position in dependence of the connection width for the 2-LW layout. Note that in the 2-LW layout the splitting surpasses the linewidth.*

In Figure 4 (b) the polarization mode splitting of the CBGs with different layouts is shown for varying connection sizes. The splitting of the 3-WG layout increases (and partly increases for the other two 3-fold layouts), while for 4-WG, 4-SW and 4-LW it stays the same or even decreases (4-SW and 4-LW). The 3-LW and 4-LW follow basically the trend of 3-SW and 4-SW, respectively. We decided to not show these datapoints for the sake of readability of the plot. At 300 nm connection width, we cannot clearly identify the mode at the 3-WG and 4-WG layouts any longer, thus there is no data included for this connection width for these layouts. The increased mode splitting for 3-WG, 3-LW and 3-SW results from the orientation of the modes and the connections. Assuming that the V-polarized mode is oriented along one connection of the innermost gap, the H-polarized mode is oriented perpendicularly and thus is not aligned with the connections. Consequently, the V-polarized mode has a bigger effective cavity length, is less confined, and thus has a longer wavelength compared to the H-polarized mode.

In addition to the previously discussed designs, we also investigated a layout with only two connections per ring (2-LW) by polarization resolved reflectivity measurements to demonstrate the influence of the connections on the polarization mode splitting. One gap to the next, the connections are turned by 90° (see Figure 1 a). By increasing the connection width, the mode splitting exceeds the cavity linewidth. Thus, the splitting can also be seen in polarization independent measurements. While both modes shift red with increasing connection size (Figure 4 c), the mode that is aligned along the innermost connections shifts significantly more. Since the effective refractive index change of the CBG is the same for both the modes, we can conclude that the dominating effect in mode shifting is the confinement change in the central disk. This behavior is analogous to the splitting observed in elliptical CBGs or micropillars [24] and offers similar application opportunities.

In general layouts with more connections per ring, e.g. 6 or 8 are possible. Layouts with 6 connections suffer from the same disadvantages as 3-WG, 3-LW and 3-SW. Layouts with 8 connections perform similar to 4 connections but not better. But since for 8 connections, particular with increasing connection width, there is not much left of the innermost gap, in our initial optimization runs we recommend 4 connections.

In conclusion, we studied seven labyrinth layouts of electrically contactable circular Bragg grating cavities that differ in position of the connections between the consecutive rings. While the configuration with a straight arrangement of connections (WG) shows a significant redshift with increasing connections, rotating the connections to long way (LW) and short way (SW) reduces this effect by increasing the confinement in the central disk. Comparing the influence of the connection width on the Q-factor illustrates that layouts with four connections have a better Q-factor than layouts with three connections at widths exceeding 100 nm. Furthermore, the FDTD simulations reveal that the 4-LW and 4-SW layouts preserve an almost Gaussian far-field pattern. In this respect, one should expect good coupling efficiency into single mode fibers. Layouts with three connections per ring increase the polarization mode splitting while layouts with four connections reduce it. We showed a splitting of over 10 nm at devices with two connections per ring. This can be used if polarized photons are required, as with elliptical devices, with the advantage of possible electrical contacts. Furthermore, this influence on the polarization mode splitting could be used to reduce unintended mode splitting, if the unintended mode splitting has a preferred direction. Such a preferred directionality could be caused for example by anisotropic etching. Also, devices with 4 connections but different connection widths along different axes can be used to achieve the same. Overall, for unpolarized device performance, we suggest the 4-LW layout for electrical contacted CBGs. Alternatively, 4-SW offers a layout with a shorter path length for the current if the device resistance is critical. For the final device, fast electronic switching must be considered. Not only resistance, but also the capacity that forms between top and bottom layer will limit switching times. A wanted maximum device resistance will result in a lower bound for the connection width. However, we showed for the 4-LW and 4-SW layouts that connection widths exceeding 300 nm are possible without decreasing the Q-factor. Thus, there is a big parameter space for the connection width to find a suitable width regarding electrical contacts.

## Acknowledgments

We want to thank the funding agencies. Especially BMBF – Project Q.LINK.X, QR.X and Qecs (Förderkennzeichen 16KIS0871, 16KISQ010 and 13N16272), and the Deutsche Forschungsgemeinschaft (DFG, German Research Foundation) within the projects HU2985/1-1 and HO5194/8-1 and the State of Bavaria for funding the research. The authors acknowledge support within the QuantERA II Program that has received funding from the European Union's Horizon 2020 research and innovation program under Grant Agreement No 101017733 (QD-E-QKD – funded through the BMBF under 16KIS1672K). Special thanks go to Silke Kuhn for processing the samples in reliable good quality and to Johannes Beierlein for helping with the first draft of the lithography layout.

## Author declarations

### Conflict of Interest

The authors have no conflicts to disclose.

### Data availability

The data that support the findings of this study are available from the corresponding author upon reasonable request.

### Intellectual property

Simon Betzold, Sven Höfling and Tobias Huber-Loyola have Patent EP21210266.9 pending.

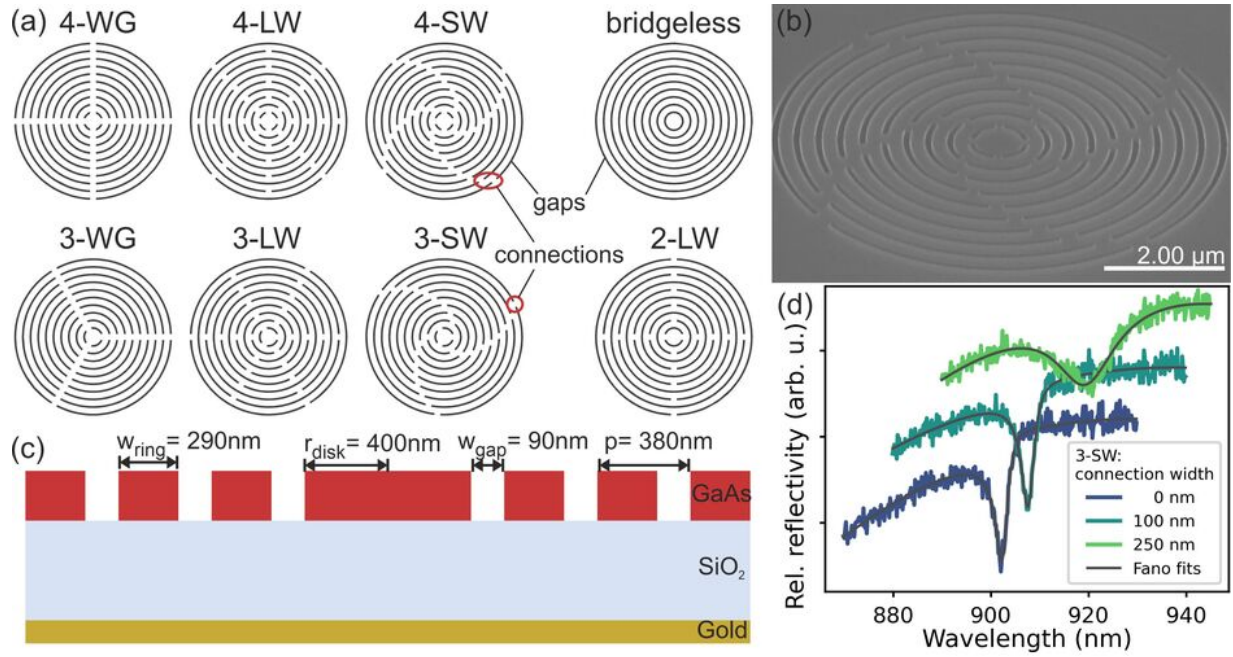

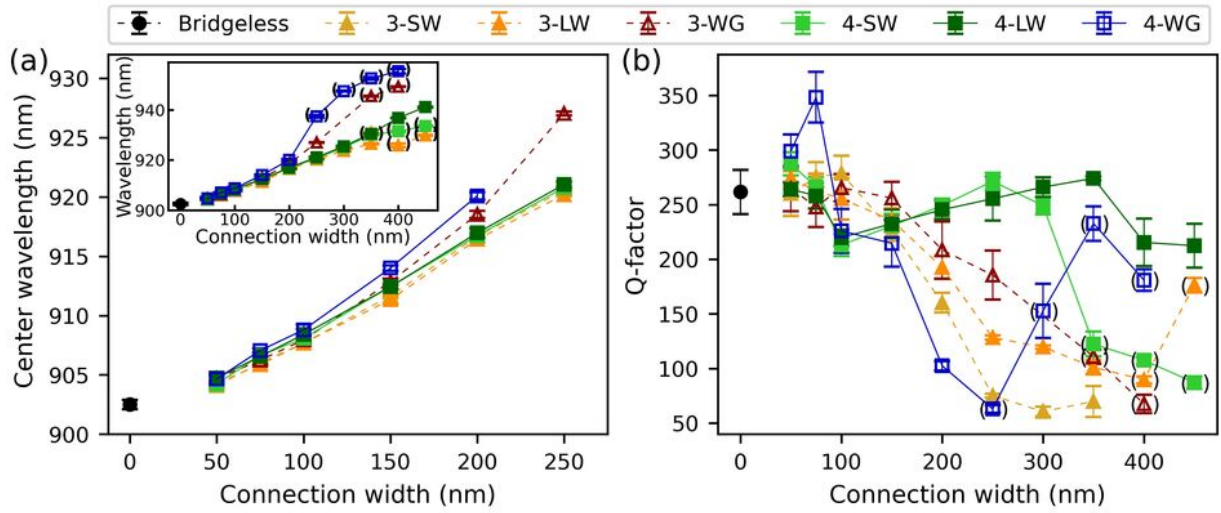

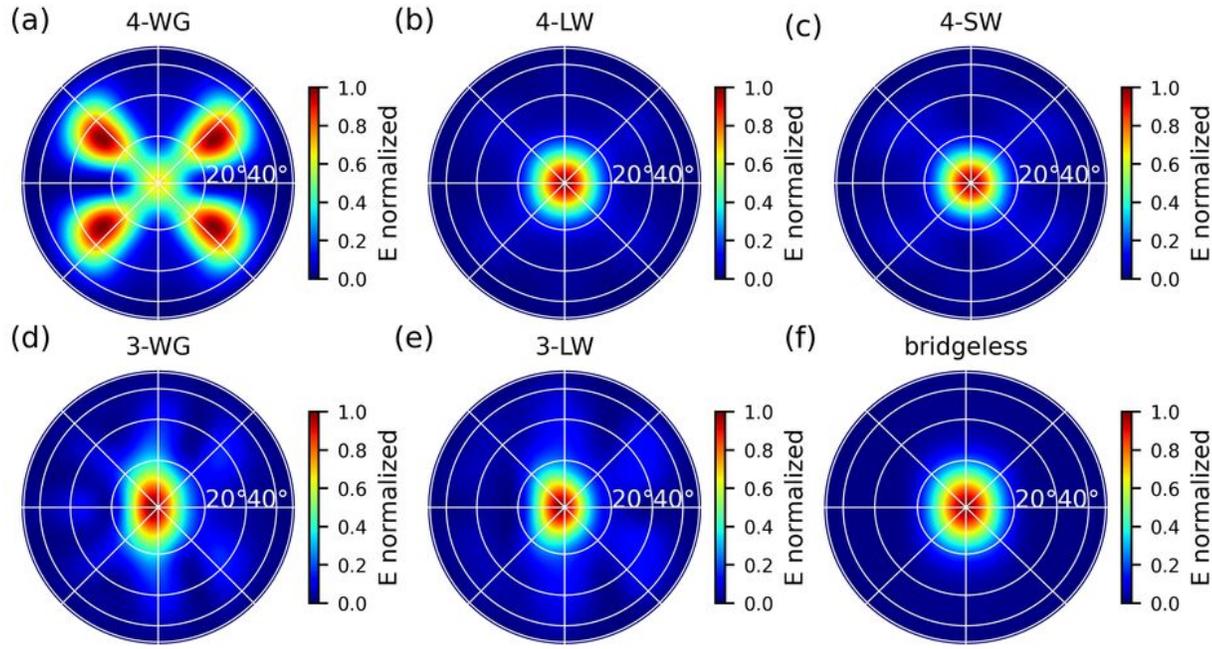

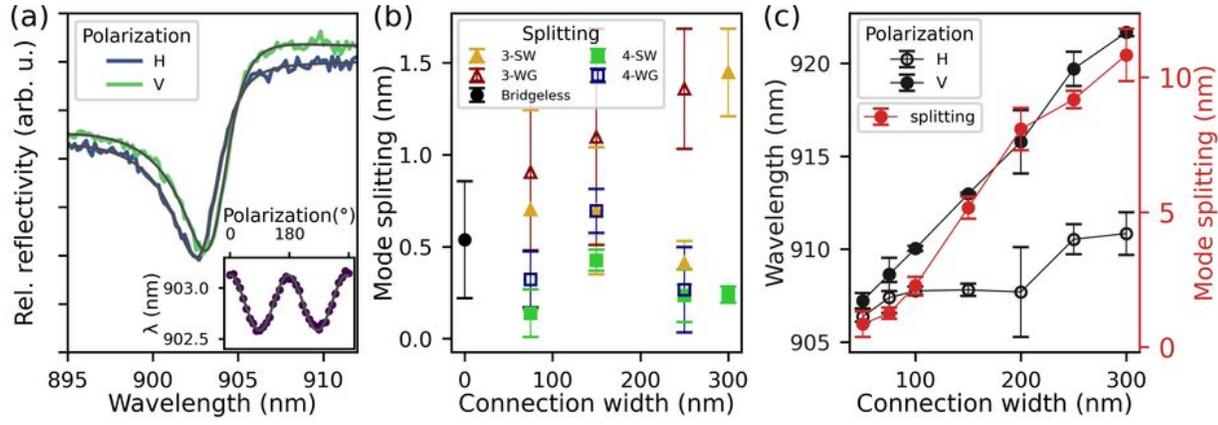